\documentstyle[twoside,fleqn,espcrc2,epsf]{article}

\newcommand{\AmS}{{\protect\the\textfont2
  A\kern-.1667em\lower.5ex\hbox{M}\kern-.125emS}}

\newcommand{\ovl}[1]{\overline{#1}}

\newcommand{\pslash}{p\kern-1ex /}
\newcommand{\Dslash}{{\cal D}\kern-1.5ex /}

\newcommand{\msbar}{{\overline {\rm MS}}}

\newcommand{\VEV}[3]{\left\langle #1\left| #2 \right| #3\right\rangle}

\title{Kaon B parameter from quenched domain-wall QCD
\thanks{Talk presented by Y.~Taniguchi}}

\author{CP-PACS Collaboration :
  A.~Ali~Khan\rlap,\address{Center for Computational Physics,
    University of Tsukuba, Tsukuba, Ibaraki 305-8577, Japan}
  S.~Aoki\rlap,\address{Institute of Physics,
    University of Tsukuba, Tsukuba, Ibaraki 305-8571, Japan}
  Y.~Aoki\rlap,$^{\rm a,b}$\thanks{address after 1 May, 2000:
        RIKEN BNL Research Center, Brookhaven National
        Laboratory, Upton, NY 11973, USA}
  R.~Burkhalter\rlap,$^{\rm a,b}$
  S.~Ejiri\rlap,$^{\rm a}$
  M.~Fukugita\rlap,\address{Institute for Cosmic Ray Research,
    University of Tokyo, Kashiwa 277-8582, Japan}
  S.~Hashimoto\rlap,\address{High Energy Accelerator Research Organization
    (KEK), Tsukuba, Ibaraki 305-0801, Japan}
  N.~Ishizuka\rlap,$^{\rm a,b}$
  Y.~Iwasaki\rlap,$^{\rm a,b}$
  T.~Izubuchi\rlap,\address{Institute of Theoretical Physics, Kanazawa
    University, Ishikawa 920-1192, Japan}
  K.~Kanaya\rlap,$^{\rm b}$
  T.~Kaneko\rlap,$^{\rm d}$
  Y.~Kuramashi\rlap,$^{\rm d}$
  K.-I.~Nagai\rlap,$^{\rm a}$
  J.~Noaki\rlap,$^{\rm a}$
  M.~Okawa\rlap,$^{\rm d}$
  H.P.~Shanahan\rlap,$^{\rm a}$\thanks{address after 15 Sept., 2000:
        Department of Biochemistry and Molecular
        Biology, University College London, London, England, UK}
  Y.~Taniguchi\rlap,$^{\rm b}$
  A.~Ukawa$^{\rm a,b}$ and
  T.~Yoshi\'e$^{\rm a,b}$
 }

\begin{document}

\begin{abstract}
We report on a calculation of $B_K$ with domain wall 
fermion action in quenched QCD. Simulations are made with a renormalization 
group improved gauge action at $\beta=2.6$ and $2.9$ corresponding to 
$a^{-1}\approx 2$GeV and $3$GeV.  Effects due to finite fifth
dimensional size $N_5$ and finite spatial size $N_\sigma$ are examined 
in detail.  Matching to the continuum operator is made perturbatively
at one loop order.  We obtain $B_K(\mu = 2 \mbox{GeV})= 0.5746(61)$,
where the error is statistical only,  
as an estimate of the continuum value in the $\msbar$ 
scheme with naive dimensional regularization. 
This value is smaller but consistent with
$B_K(\mu = 2 \mbox{GeV})= 0.628(42)$ obtained by the JLQCD Collaboration
 using the Kogut-Susskind quark action.
Results for strange quark mass are also reported.
\end{abstract}

\maketitle

\section{Introduction}

The Kaon B parameter $B_K$ is an important quantity to pin down 
the CKM matrix, thereby gaining an understanding on CP violation 
in the Standard Model.

A crucial ingredient in a precision calculation of $B_K$
is chiral symmetry.  For this reason, 
the best result so far has been obtained \cite{jlqcd} 
with the Kogut-Susskind (KS) quark action, which keeps $U(1)$ subgroup 
of chiral symmetry at finite lattice spacings.  
Carrying out a systematic and extensive set of simulations 
to control ${\cal O}(a^2)$ scaling violation and 
${\cal O}(\alpha_{\rm \ovl{MS}}^2)$ errors that arise with the use of one-loop 
perturbative renormalization factors, the value 
$B_K(\mu=2{\rm GeV})=0.628(42)$ was obtained in the continuum limit 
in the $\msbar$ scheme with naive dimensional regularization (NDR). 

The Wilson fermion action has a problem of explicit
chiral symmetry breaking, which causes a nontrivial mixing between
four-quark operators with different chiralities.
This problem has been treated with several non-perturbative
renormalization methods \cite{GBS1997,jlqcd-wilson,Allton1999,UKQCD1999}.
The resultant values of $B_K$ are consistent with that of the KS action, 
but the numerical error is large.

The domain wall fermion formalism offers a possibility of a calculation 
preserving full chiral symmetry \cite{blum-soni}.  
Here we report a summary of a quenched calculation of $B_K$
with the Shamir's variant \cite{Shamir93} of the formalism and   
a renormalization group (RG) improved gauge action \cite{Iwasaki83}. 
The latter choice is motivated by our result \cite{cppacs-dwf} 
that chiral symmetry is much better realized with this action than 
with the plaquette gauge action. 
We may also expect that scaling behavior 
of $B_K$ is improved with the use of the RG-improved action. 
We also report on the strange quark mass obtained from meson mass measurements
in our simulation.   

\section{Numerical simulation}

\begin{table}[tb]
\setlength{\tabcolsep}{0.6pc}
\begin{center}
\caption{
Simulation parameters together with the number of configurations 
analyzed shown in bold letters. 
}
\begin{tabular}{cccccc}
\hline
 & \multicolumn{3}{c}{$\beta=2.6$} & \multicolumn{2}{c}{$\beta=2.9$}\\
\hline
$a^{-1}$(GeV)  & \multicolumn{3}{c}{1.81(4)} & \multicolumn{2}{c}{2.87(7)} \\
\hline
$N_t$  & \multicolumn{3}{c}{40} & \multicolumn{2}{c}{60} \\
$N_\sigma$ & 16 & 24 & 32 & 24 & 32 \\
\hline
$N_5=16$ & {\bf122} & {\bf 76} & {\bf 25} & {\bf 76} & {\bf 50} \\
$N_5=32$ & $-$   & {\bf 50} & $-$ & $-$ & $-$ \\
\hline
\end{tabular}
\label{tab:config}
\end{center}
\vspace*{-6ex}
\end{table}

Simulations are made at two values of coupling,  
$\beta=2.6$ and $2.9$ corresponding 
to a lattice spacing $a^{-1}=1.81(4)$ GeV and $2.87(7)$ GeV
determined by $\rho$ meson mass,
in order to check the scaling behavior.

The choice of lattice size $N_\sigma^3 \times N_t \times N_5$ is 
made as follows: 
(i) Our main runs use $24^3\times 40\times 16$ at $\beta=2.6$, 
and $32^3\times 60\times 16$ at $\beta=2.9$.  
(ii) Spatial size dependence is examined at $\beta=2.6$
varying the spatial size from $N_\sigma=24$ to $N_\sigma=16$ and 32:
They correspond to the physical size $aN_\sigma\sim1.7, 2.6, 3.4$ fm.
(iii) Dependence on the fifth dimensional length $N_5$ is also examined 
at $\beta=2.6, N_\sigma=24$ using $N_5=16$ and $N_5=32$.

We choose $m_f=0.01, 0.02, 0.03, 0.04$ for the bare quark mass,
covering the range $m_{PS}/m_V$= 0.4 -- 0.8.
The {\it u-d} and {\it s} quarks are assumed degenerate in the 
analysis for $B_K$. The domain wall height is taken to be $M=1.8$.
The number of configurations 
analyzed is given in Table \ref{tab:config}.

We measure the Kaon B parameter,  
\begin{eqnarray}
B_K =
\frac{
\VEV{K}{\ovl{s}\gamma_\mu(1-\gamma_5)d \ovl{s}\gamma_\mu(1-\gamma_5)d}{K}}
{\frac{8}{3}\VEV{K}{\ovl{s}\gamma_\mu\gamma_5d}{0}
 \VEV{0}{\ovl{s}\gamma_\mu\gamma_5d}{K}},
\end{eqnarray}
and the matrix element divided by the pseudo scalar density, 
\begin{eqnarray}
B_P =
\frac{
\VEV{K}{\ovl{s}\gamma_\mu(1-\gamma_5)d \ovl{s}\gamma_\mu(1-\gamma_5)d}{K}}
{\VEV{K}{\ovl{s}\gamma_5d}{0} \VEV{0}{\ovl{s}\gamma_5d}{K}}
\end{eqnarray}
which should vanish at $m_\pi\to 0$.  The extraction of these quantities 
from Kaon Green functions follows the standard procedure (see, {\it e.g.,} 
Ref.~\cite{jlqcd}) using the Dirichlet boundary condition in time 
and wall sources for quark propagators. 
We simultaneously evaluate hadron masses.
The physical point for quark masses is calculated by linearly 
fitting the light hadron masses 
$m_{PS}^2$ and $m_V$
as a function of $m_f$,
and using the experimental values of $m_\pi/m_\rho$, $m_K/m_\rho$ 
as input.

\section{Operator matching}

We carry out matching of the lattice and continuum operators 
at a scale $q^*=1/a$ using one-loop perturbation theory \cite{zfactor} and 
the $\msbar$ scheme with NDR in the continuum. 
In the domain wall formalism the renormalization factor of an $n$-quark 
operator $O_n$ has a generic form
\begin{equation}
Z=(1-w_0^2)^{n/2}Z_w^{n/2}Z_{O_n}
\end{equation}
where $w_0=1-M$ and $Z_{w,{O_n}}=1+O(g^2)$.  We apply tadpole improvement 
by explicitly moving the one-loop correction to the domain wall height 
$M$ from $Z_w$ to $w_0$, and by factoring out a tadpole factor 
$u^{n/2}=P^{n/8}$ 
with $P$ the plaquette from $Z_{O_n}$. A mean-field estimate appropriate for 
the RG-improved action is used for calculating the coupling constant 
$g^2_{\overline{\rm MS}}(\mu)$ \cite{cppacs-full}.       
The continuum value at a physical scale {\it e.g.,} $\mu=2.0$ GeV, is 
obtained {\it via} a renormalization group running 
from $q^*=1/a$ to $\mu$.

For $B_K$ the factor $(1-w_0^2)^2Z_w^2$ cancels out, and 
the one-loop value given by the ratio
\begin{equation}
\frac{Z_{O_4}}{Z_A^2}=\frac{1-(4\log q^*a+13.6)g^2/(16\pi^2)}
{(1-6.26g^2/(16\pi^2))^2}
\end{equation}
turned out to be very near unity {\it i.e.,} 
$Z_{B_K}^{\ovl{\rm MS}}(q^*=1/a)=0.984(\beta=2.6)$ and 
$0.988(\beta=2.9)$.
The $Z$ factor at the scale $\mu=2$GeV obtained with a 2-loop 
running becomes
$Z_{B_K}^{\ovl{\rm MS}}(\mu=2{\rm GeV})=0.979(\beta=2.6)$ and 
$1.006(\beta=2.9)$.

\section{Results for $B_K$}

\subsection{Chiral property}

\begin{figure}[tb]
\centerline{\epsfxsize=13pc \epsfbox{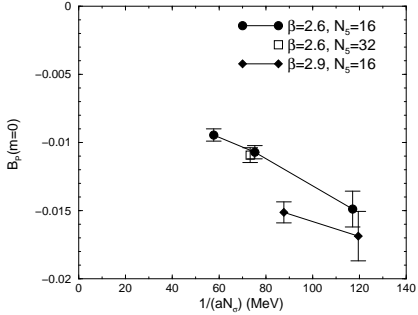}}
\vspace*{-7ex}
\caption{The matrix element $B_P$ at $m_f=0$ as a function of spatial size.}
\label{fig:BP-V}
\end{figure}

We start by investigating the chiral property of the system through $B_P$.
The renormalized value of $B_P$, extrapolated to $m_f=0$ linearly in $m_f$, 
is plotted as a function of the inverse of spatial size in
Fig.~\ref{fig:BP-V}.

Since results for the fifth dimensional size $N_5=16$ (filled circle)
and 32 (open square) agree with each other, the observed non-zero value, 
albeit very small, should be due to a finite spatial-size
effect. The decrease of the magnitude of $B_P$ toward larger spatial sizes
supports this interpretation.
The existence of finite-size effect is consistent with the result for 
the pion mass in Ref.~\cite{cppacs-dwf}, where a non-zero pion mass 
of a magnitude expected from finite-size effects observed for the KS fermion 
was found for $N_5\ge16$ at $\beta=2.6$.
It is a future problem to see whether and how values of $B_P$ vanish toward
the infinite volume limit.

\subsection{$B_K$}

\begin{figure}[tb]
\vspace*{-4ex}
\centerline{\epsfxsize=12pc \epsfbox{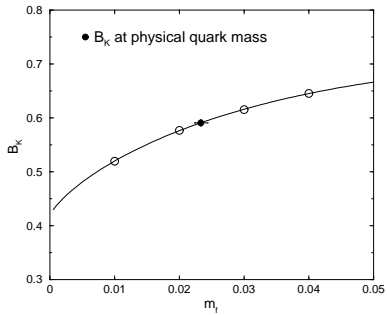}}
\vspace*{-7ex}
\caption{Bare $B_K$ interpolated as a function of 
$m_fa$ at $\beta=2.6$ for a $24^3\times40\times16$ lattice.}
\label{fig:BK-mf}
\vspace*{-4.5ex}
\end{figure}

\begin{figure}[tb]
\centerline{\epsfxsize=12pc \epsfbox{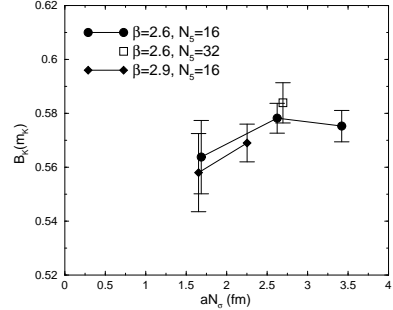}}
\vspace*{-7ex}
\caption{Renormalized $B_K$ as a function of spatial size.}
\label{fig:BK-V}
\end{figure}

\begin{figure}[tb]
\vspace*{-4ex}
\centerline{\epsfxsize=12pc \epsfbox{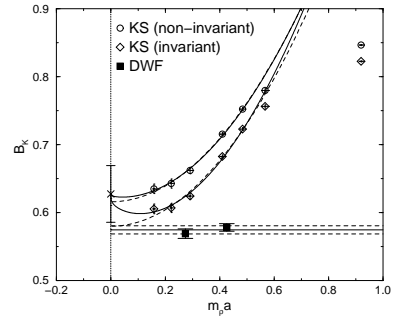}}
\vspace*{-7ex}
\caption{Scaling behavior of $B_K$.
Previous results with the KS action\cite{jlqcd} are also shown.
}
\label{fig:BK-a}
\vspace*{-5ex}
\end{figure}

The bare value of $B_K$ is interpolated as a function of $m_fa$ with the 
formula given by chiral perturbation theory,
\begin{eqnarray}
B_K=B\left(1-3c (m_fa)\log(m_fa) + b(m_fa)\right)
\end{eqnarray}
as shown in Fig.~\ref{fig:BK-mf}. 
The physical value of $B_K$ is obtained at half the strange quark mass
$m_sa/2$ (solid circle in Fig.~\ref{fig:BK-mf}) 
which is estimated from the experimental value of $m_K/m_\rho$.

We plot the renormalized value of $B_K({\rm NDR};\mu=2 {\rm GeV})$
as a function of the spatial size in Fig.~\ref{fig:BK-V}.
Filled circles and diamonds are data at $\beta=2.6$ and $2.9$ 
keeping the same fifth dimensional size $N_5=16$.
The spatial size dependence is mild especially for 
$N_\sigma a \ge 2.6$ fm at $\beta=2.6$. 
This result is consistent with that of a previous finite spatial size study 
with the staggered quark action \cite{jlqcd}.
We conclude that the size of about 2.4 fm 
used in our main runs is
sufficient to avoid spatial size effects for $B_K$.

In Fig.~\ref{fig:BK-V} an open square shows the result for $N_5=32$ 
on a $24^3\times40$ four-dimensional lattice at $\beta=2.6$. 
Since this data is consistent with that at $N_5=16$ within the statistical
error, the fifth dimensional size of $N_5=16$ is sufficient for $B_K$. 

Our final results from the main runs are shown in Fig.~\ref{fig:BK-a}.
We observe a good scaling behavior, and making a 
constant extrapolation $B_K(a)=B_K$, we find 
$B_K(\mu=2\mbox{GeV})= 0.5746(61)$ where the error is statistical only.
The open symbols and the associated lines represent results from a 
previous calculation with the staggered quark action \cite{jlqcd}, where
gauge invariant and non-invariant four-quark operators are used.
Compared to the result in the continuum limit $B_K=0.628(42)$ from this
work, our present result is smaller but consistent within the error. 

\section{Strange quark mass}

\begin{figure}[tb]
\centerline{\epsfxsize=12pc \epsfbox{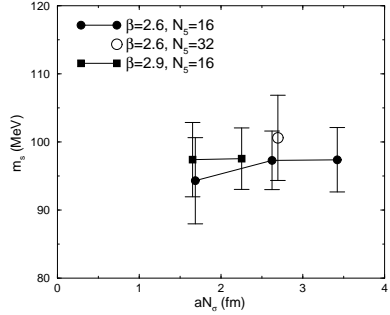}}
\vspace*{-7ex}
\caption{Renormalized strange quark mass as a function of spatial size.}
\label{fig:ms-V}
\vspace*{-5ex}
\end{figure}

We show results for the strange quark mass in the $\overline{\rm MS}$ scheme 
at 
$\mu=2$ GeV in Fig.~\ref{fig:ms-V}, 
where $m_K$ is used as input.
We observe that the fifth dimensional size $N_5=16$ and spatial size
$N_\sigma a=$ $2.4$ fm 
is sufficient for a reliable determination of 
strange quark mass, giving $m_s \approx$ 97 MeV.
A recent result $m_s =$ 110(2)(22) MeV \cite{RikenBNL} 
using the domain wall fermions but with the plaquette gauge action  
at $\beta=$ 6.0 is also consistent with ours, if 
the systematic error from the conversion to the $\msbar$ scheme
is taken into account.

\begin{figure}[tb]
\centerline{\epsfxsize=12pc \epsfbox{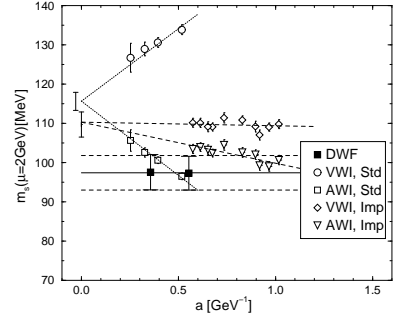}}
\vspace*{-7ex}
\caption{
Scaling behavior of the strange quark mass compared with those from 
the standard (Std) \cite{cppacs-quenched} and 
an improved (Imp)\cite{cppacs-full} 4-d actions.
VWI and AWI represent vector and axial-vector Ward-identity masses,
respectively.
}
\label{fig:ms-a}
\vspace*{-5ex}
\end{figure}

Our results exhibit a good scaling behavior as shown in Fig.~\ref{fig:ms-a}.
Making a constant fit yields $m_s^{\msbar}(2 {\rm GeV}) = 97.4(4.4)$ MeV. 
Compared with results from the conventional 
4-dimensional formalisms (the plaquette and Wilson quark 
action \cite{cppacs-quenched} and an RG-improved gauge action and 
the clover quark action \cite{cppacs-full}),  
our result is 3--4$\sigma$ (13--18~MeV) smaller.
Reduction of systematic errors from one-loop renormalization factors and 
finite-size effects would be needed to see if there is an actual disagreement. 

This work is supported in part by Grants-in-Aid
of the Ministry of Education (Nos.
10640246, 10640248, 10740107, 11640250, 11640294, 11740162,
12014202, 12304011, 12640253, 12740133).
SE, TK, KN, JN
and HPS are JSPS Research Fellows.
AAK is supported by JSPS Research for the Future Program
(No. JSPS-RFTF 97P01102).

\newcommand{\J}[4]{{#1} {#2} (#3) #4}
\newcommand{\AP}{Ann.~Phys.}
\newcommand{\CMP}{Commun.~Math.~Phys.}
\newcommand{\IJMP}{Int.~J.~Mod.~Phys.}
\newcommand{\MPL}{Mod.~Phys.~Lett.}
\newcommand{\NP}{Nucl.~Phys.}
\newcommand{\NPSup}{Nucl.~Phys.~B (Proc.~Suppl.)}
\newcommand{\PL}{Phys.~Lett.}
\newcommand{\PR}{Phys.~Rev.}
\newcommand{\PRL}{Phys.~Rev.~Lett.}
\newcommand{\PTP}{Prog. Theor. Phys.}
\newcommand{\Suppl}{Prog. Theor. Phys. Suppl.}

\end{document}